# From Slaves to Synths? Superintelligence and the Evolution of Legal Personality

Simon Chesterman*

*This chapter examines the evolving concept of legal personality through the lens of recent developments in artificial intelligence and the possible emergence of superintelligence. Legal systems have long been open to extending personhood to non-human entities, most prominently corporations, for instrumental or inherent reasons. Instrumental rationales emphasize accountability and administrative efficiency, whereas inherent ones appeal to moral worth and autonomy. Neither is yet sufficient to justify conferring personhood on AI. Nevertheless, the acceleration of technological autonomy may lead us to reconsider how law conceptualizes agency and responsibility. Drawing on comparative jurisprudence, corporate theory, and the emerging literature on AI governance, this chapter argues that existing frameworks can address short-term accountability gaps, but the eventual development of superintelligence may force a paradigmatic shift in our understanding of law itself. In such a speculative future, legal personality may depend less on the cognitive sophistication of machines than on humanity's ability to preserve our own moral and institutional sovereignty.*

---

* David Marshall Professor of Law, National University of Singapore. This chapter draws on material first published as Simon Chesterman, 'Artificial Intelligence and the Limits of Legal Personality' (2020) 69 *International and Comparative Law Quarterly* 819; Simon Chesterman, *We, the Robots? Regulating Artificial Intelligence and the Limits of the Law* (Cambridge University Press 2021); and Simon Chesterman, 'I, Robot? Legal Personality for Robots and the Android Fallacy' in Woodrow Barfield, Ugo Pagallo, and Yueh-Hsuan Weng (eds), *The Cambridge Handbook of the Law, Policy, and Regulation for Human-Robot Interaction* (Cambridge University Press 2024) 173.

# 1 Introduction: Thinking Machines

The question of whether artificial systems might one day qualify as legal persons — entities capable of holding rights and duties before the law — has moved from speculative fiction to a matter of practical jurisprudence. For much of the late twentieth century, digital technologies were understood as tools: accelerants of human agency rather than autonomous agents in their own right. The so-called 'information superhighway' merely provided a faster route for human and corporate interaction. Yet as algorithmic systems acquired speed, autonomy, and opacity, the assumption that human beings must always stand behind digital acts became harder to sustain. When a trading algorithm can wipe a trillion dollars from markets in minutes,[1] or when a self-driving car kills a pedestrian whose death cannot easily be attributed to a human decision, the familiar architecture of accountability begins to buckle.[2]

The fascination with non-human agents is hardly new. From Mary Shelley's *Frankenstein* (1818) to Karel Čapek's *R.U.R. (Rossum's Universal Robots)* (1920), the cultural imagination has long explored humanity's ambivalence toward artificial life. Isaac Asimov's 'Three Laws of Robotics' distilled this anxiety into a formula: machines should obey, protect, and never harm humans.[3] Alan Turing reframed intelligence as imitation rather than agency; his 1950 'imitation game' proposed that if a computer could convincingly mimic a human in conversation, we might properly call it intelligent.[4] The early program 'Eliza' offered a first taste of this illusion. Users who conversed with the text-based 'therapist' in the 1960s often believed that the machine understood them — even after being told it merely rephrased their statements.[5] Turing's insight was prophetic: people are willing to attribute humanity to pattern-recognizing machines that reflect their own language and emotions, identifying agency and moral worth in a simulacrum of themselves.

---

[1] Andrei Kirilenko et al, 'The Flash Crash: High-Frequency Trading in an Electronic Market' (2017) 72 Journal of Finance 967.

[2] Visa A.J. Kurki, *A Theory of Legal Personhood* (Oxford University Press 2019); Ryan Abbott, *The Reasonable Robot: Artificial Intelligence and the Law* (Cambridge University Press 2020).

[3] Isaac Asimov, 'Runaround', *Astounding Science Fiction* (March 1942) 94.

[4] A.M. Turing, 'Computing Machinery and Intelligence' (1950) 59 Mind 433.

Half a century later, generative AI has completed the transition from mechanical response to synthetic expression. Large language models and their ilk now compose essays, produce art, and simulate empathy with disarming fluency and conviction.[6] The boundaries between tool, agent, and companion have blurred. Popular culture mirrors this unease: films such as 'Her' and 'Ex Machina' imagine intimate relationships between humans and code; corporations market 'social robots' designed to comfort the elderly or tutor children,[7] or as erotic soulmates.[8] These developments foreground legal questions previously of interest only to philosophers and legal theorists: when artificial entities act in the world, on what basis — if any — should they be recognized as persons under the law?

To be sure, contemporary law already accommodates entities that are not human. Corporations, foundations, even ecosystems have been granted juridical personality, enabling them to own property, enter contracts, and be held liable for wrongs. The rationales for such recognition fall broadly into two categories. The instrumental rationale treats personhood as a convenience: a legal fiction that simplifies responsibility and commercial interaction. The inherent rationale, by contrast, links personhood to moral worth — the recognition of an entity as an end in itself.[9] These two strands intertwine in debates about artificial intelligence. Some advocates argue that conferring legal personality on AI would close accountability gaps, ensuring that autonomous systems capable of harm can also bear liability.[10] Others contend that once AI becomes indistinguishable from humans in reasoning and communication —

---

[5] Richard S. Wallace, 'The Anatomy of ALICE' in Robert Epstein, Gary Roberts, and Grace Beber (eds), *Parsing the Turing Test: Philosophical and Methodological Issues in the Quest for the Thinking Computer* (Springer 2009) 181.

[6] Simon Chesterman, 'Good Models Borrow, Great Models Steal: Intellectual Property Rights and Generative AI' (2025) 44(1) Policy and Society 23.

[7] Belinda J. Dunstan et al (eds), *Cultural Robotics: Social Robots and Their Emergent Cultural Ecologies* (Springer 2023).

[8] Michael J. Quinn and Jeff Riley, 'Companion Robots: A Debate' (2024) 2024(December) Ubiquity 1.

[9] Ngaire Naffine, 'Who Are Law's Persons? From Cheshire Cats to Responsible Subjects' (2003) 66 Modern Law Review 346.

[10] Joanna J. Bryson, Mihailis E. Diamantis, and Thomas D. Grant, 'Of, for, and by the People: The Legal Lacuna of Synthetic Persons' (2017) 25 Artificial Intelligence and Law 273.

once it 'passes' Turing's test — it would be unjust to *deny* it corresponding moral and legal status.[11]

Until recently, such arguments were speculative. Then came a series of gestures that, however symbolic, signalled growing unease with human exclusivity. In 2017 Saudi Arabia conferred 'citizenship' on Sophia, a humanoid robot developed by Hanson Robotics, while Tokyo granted online 'residency' to a chat-bot representing a seven-year-old boy.[12] The same year, the European Parliament urged its Commission to consider a 'specific legal status for robots,' suggesting that the most advanced systems might one day be treated as 'electronic persons' responsible for damage they cause, and possibly granting them powers to take decisions and interact with third parties.[13] These were publicity stunts rather than jurisprudential revolutions — Sophia was, in essence, a sophisticated chatbot with a silicone face[14] — but they exposed the tension between technological capability and legal imagination.

Law's impulse to assign responsibility wherever harm occurs is both a strength and a limitation. When autonomous systems act in unpredictable ways, identifying a culpable human or corporate agent becomes difficult. Was the fault in the programmer's code, the operator's oversight, the manufacturer's design, or the algorithm's independent learning? Traditional doctrines of negligence and vicarious liability presume a chain of human causation. As that chain weakens, policymakers search for substitutes — potentially including the creation of new 'electronic persons'. Yet, this approach may widen the very accountability gaps it seeks to close by allowing corporations to externalize blame onto synthetic agents of their own making.

Behind these pragmatic debates lies a more profound inquiry: what makes something a 'person' at all? The history of law is also a history of expanding personhood. Slaves, women, and indigenous peoples were once treated as property or dependants; over centuries, they were recognized as rights-bearing subjects through moral and political struggle.[15] The long arc of the moral universe, as Dr Martin Luther King Jr. reminded us, bends toward justice — but it bends slowly and often selectively. Extending that arc to non-humans — animals, ecosystems, or machines — raises the question of whether personhood tracks moral worth, sentience, or simply the capacity to participate usefully in the legal order.

---

[11] David J. Gunkel, *Robot Rights* (MIT Press 2018); Vincent C. Müller, 'Is It Time for Robot Rights? Moral Status in Artifcial Entities' (2021) 23 Ethics and Information Technology 579.

[12] Olivia Cuthbert, 'Saudi Arabia Becomes First Country to Grant Citizenship to a Robot', *Arab News* (26 October 2017); Anthony Cuthbertson, 'Artificial Intelligence "Boy" Shibuya Mirai Becomes World's First AI Bot to Be Granted Residency', *Newsweek* (6 November 2017).

[13] European Parliament Resolution with Recommendations to the Commission on Civil Law Rules on Robotics (2015/2103(INL)) (European Parliament, 16 February 2017),

[14] Dave Gershgorn, 'Inside the Mechanical Brain of the World's First Robot Citizen', *Quartz* (12 November 2017).

The analogy between historical emancipation and machine recognition is powerful but perilous. Equating AI with oppressed humans risks trivializing the moral achievement of human rights while overlooking the central distinction between sentience and simulation. As Neil Richards and William Smart observe, the 'android fallacy' tempts us to treat any system that behaves socially as possessing moral agency.[16] But legal personality built on anthropomorphic projection would rest on illusion, not ethics or law.

This chapter examines how the concept of legal personality might evolve under the pressure of artificial intelligence and, ultimately, superintelligence. Section two revisits the instrumental and inherent rationales for extending personality, drawing lessons from corporate law and animal rights advocacy to assess whether AI merits similar treatment. Section three explores the future possibility of superintelligence — entities surpassing human cognition in every domain — and considers the range of legal and ethical responses, from precautionary prohibition to normative alignment. The conclusion reflects on what these debates reveal about humanity's own conception of law, responsibility, and moral community.

The working hypothesis is deliberately conservative. Most legal systems presently *could* confer personality on AI, but none *must* — and, I would argue, all presently *should* refrain from doing so. Taking such a step today would obscure rather than resolve the core issues of accountability and moral status. The challenge is not only whether machines should enter the circle of legal persons, but whether, if and when they do, humanity will still control the pen that draws that circle. As technology accelerates, the question may ultimately reverse: not whether we recognize AI under law, but whether they will recognize us.

---

[15] Lee Holcombe, *Wives and Property: Reform of the Married Women's Property Law in Nineteenth-Century England* (Martin Robertson 1983); Jean Allain (ed), *The Legal Understanding of Slavery: From the Historical to the Contemporary* (Oxford University Press 2013).

[16] Neil M. Richards and William D. Smart, 'How Should the Law Think About Robots?' in Ryan Calo, A. Michael Froomkin, and Ian Kerr (eds), *Robot Law* (Edward Elgar 2016) 3 at 18-21.

## 2  Instrumental and Inherent Personality

Legal personality is the skeleton upon which every legal system is built. The question of who can act, who can be the subject of rights and duties, is a precursor to almost every other issue. Yet close examination of these foundations reveals surprising uncertainty and disagreement. Despite this, as John Dewey observed a century ago, 'courts and legislators do their work without such agreement, sometimes without any conception or theory at all' regarding the nature of personality. Indeed, he went on, recourse to theory has 'more than once operated to hinder rather than facilitate the adjudication of a special question of right or obligation'.[17]

In practice, the vast majority of legal systems recognize two forms of legal person: natural and juridical. Natural persons are recognized because of the simple fact of being human.[18] Juridical persons, by contrast, are non-human entities that are granted certain rights and duties by law. Corporations and other forms of business associations are the most common examples, but many other forms are possible. Religious, governmental, and intergovernmental entities may also act as legal persons at the national and international level.

It is telling that these are all aggregations of human actors, though there are examples of truly non-human entities being granted personhood. In addition to the examples mentioned in the introduction, these include temples in India, a river in New Zealand, and the entire ecosystem of Ecuador. There seems little question that a state *could* attribute some kind of personality to new entities like AI systems; if that happens, recognition would likely be accorded by other states also.[19]

Scholars and law reform bodies have already proposed attributing AI systems with some form of legal personality to help address liability questions, such as an automated driving system entity in the case of driverless cars whose behavior may not be under the control of their 'drivers' or predictable by their manufacturers or owners.[20] A few writers have gone further, arguing that procedures need to be put in place to try robot criminals, with provision for 'punishment' through reprogramming or, in extreme cases, destruction.[21]

---

[17] John Dewey, 'The Historic Background of Corporate Legal Personality' (1926) 35 Yale Law Journal 655, 660.

[18] Naffine (n 9).

[19] Kurki (n 2) 127-52.

[20] Simon Chesterman, 'Artificial Intelligence and the Problem of Autonomy' (2020) 1 Notre Dame Journal on Emerging Technologies 210.

## 2.1 Bodies Corporate

These arguments suggest an instrumental approach to personality, but scholarly explanations of the most common form of juridical person — the corporation — offer disparate justifications for its status as a separate legal person that help answer the question of whether that status should be extended to AI systems also. From its earliest incarnations in medieval guilds and ecclesiastical institutions to the joint-stock companies of the seventeenth century, personality served pragmatic functions: enabling continuity and limiting risk.[22] Competing theories developed to explain the fiction of personality layered on top of these functional goals.

The aggregate or contractarian theory views the corporation as a collection of natural persons who agree to act as one for convenience.[23] On this view, the 'person' adds no metaphysical substance; it is merely shorthand for a web of private arrangements. The fiction and concession theories emphasize state creation. The corporation is an 'artificial being, invisible, intangible, and existing only in contemplation of law'.[24] Its existence reflects legislative will, not social fact.

The realist theory, by contrast, contends that corporations attain a measure of independent reality. They act through their agents but also generate autonomous interests and identities. At its most extreme, it is argued that corporations are not only legal but also *moral* persons.[25] This approach is favoured more by theorists and sociologists than legislators and judges, but echoes the tension highlighted in the introduction to this chapter: that legal personality is not merely bestowed but *deserved*. In practice, however, actual recognition as a person before the law remains in the gift of the state.[26]

---

[21] Christina Mulligan, 'Revenge Against Robots' (2018) 69 South Carolina Law Review 579; Ying Hu, 'Robot Criminals' (2019) 52 University of Michigan Journal of Law Reform 487, 503-07.

[22] Christine E. Amsler, Robin L. Bartlett, and Craig J. Bolton, 'Thoughts of Some British Economists on Early Limited Liability and Corporate Legislation' (1981) 13 History of Political Economy 774; Giuseppe Dari-Mattiacci et al, 'The Emergence of the Corporate Form' (2017) 33 Journal of Law, Economics, and Organization 193.

[23] Ronald Coase, 'The Nature of the Firm' (1937) 4 Economica 386.

[24] *Trustees of Dartmouth Coll. v. Woodward*, 17 US 518, 636 (1819).

[25] Peter French, 'The Corporation as a Moral Person' (1979) 16 American Philosophical Quarterly 207.

Each of these theories has been invoked in contemporary debates over artificial intelligence. If corporations — once seen as abstractions — could be recognized as persons to facilitate commerce, why not intelligent machines to facilitate accountability? Advocates of 'electronic personhood' argue that as AI systems assume tasks involving independent judgment and adaptive learning, they should be recognized as legal actors capable of rights and obligations.

Yet the analogy fails in crucial respects. Corporations are aggregations of humans; their decisions can ultimately be traced to human agents and shareholders. AI systems, by contrast, are human-made artifacts whose actions may be opaque even to their creators. Extending personhood to such entities risks reversing accountability: rather than closing gaps, it could enable human actors to hide behind synthetic fronts. The European Parliament's suggestion that 'electronic persons' might bear liability for harm illustrates the danger. If an autonomous vehicle were its own legal person, its manufacturer could argue that damages should be paid by the machine itself — an empty formality if that 'person' owns nothing.[27]

Corporate law already struggles with the moral hazards of limited liability. The 'corporate veil' can be pierced when used to perpetrate fraud or injustice.[28] But extending a similar shield to AI would compound the problem. Instead of shareholders, there would be engineers and developers who program autonomy into systems and then disclaim control. The resulting diffusion of responsibility would not modernize the law; it would hollow it out.

Nor is new personhood required to enable AI participation in commerce. Electronic agents already conclude trades and negotiate digital contracts on behalf of humans. The Uniform Electronic Transactions Act and similar frameworks validate such actions without recognizing machines as persons. When an algorithm errs, the law treats it as an extension of its user or principal, just as a pen's mis-stroke binds the author, not the pen.

---

[26] Katsuhito Iwai, 'Persons, Things and Corporations: The Corporate Personality Controversy and Comparative Corporate Governance' (1999) 47 American Journal of Comparative Law 583; Susan Mary Watson, 'The Corporate Legal Person' (2019) 19 Journal of Corporate Law Studies 137.

[27] Bryson, Diamantis, and Grant (n 10) 287.

[28] David Millon, 'Piercing the Corporate Veil, Financial Responsibility, and the Limits of Limited Liability' (2007) 56 Emory Law Journal 1305.

The logic of instrumental personality is therefore double-edged. It can simplify transactions and clarify responsibility, but it can also be exploited to obscure it. AI challenges us to ensure that the tool remains tethered to its maker. Creating new legal 'entities' may satisfy a conceptual neatness — the sense that every act requires an actor — but it risks producing responsibility without retribution, a personhood without persons. As Dewey warned, theoretical elegance should never impede 'the adjudication of a special question of right or obligation'.[29]

In the near term, the more practical course is to strengthen existing doctrines of liability and agency: ensure traceability of code, impose strict duties of oversight, and require financial security for potential harms. If needed, corporate structures could be created for AI systems, though any limitation on liability should be transparent and subject to the rules that have bound other corporate actors for decades. These approaches preserve human accountability while recognizing that law can adapt without inventing new metaphysical beings.

## 2.2  Beings Like Us

If instrumental personality answers the question *who can act*, inherent personality asks *who ought to matter*. It is the moral dimension of legal recognition — the acknowledgment that some entities deserve protection not because it is useful but because it is right.

Historically, this category has expanded unevenly. Roman law recognized only the free male citizen as a legal subject; slaves were property.[30] Under English common law, Blackstone could still write that 'husband and wife are one person, and the husband is that person'.[31] The long struggle for the recognition of women, indigenous peoples, and minorities as full persons reveals that personhood is not a static status but a political achievement. Its expansion reflects evolving moral sentiment, not ontological discovery.

Interestingly, some arguments in favour of legal personality for AI draw not on this progressivist narrative of natural personhood but on the darker history of slavery. Andrew Katz and Ugo Pagallo, for example, find analogies with the ancient Roman law mechanism of *peculium*, whereby a slave lacked legal personality and yet could operate as more than a mere agent for his master.[32] (In 2017, a digital bank of that name was established in France — presumably for investors who never studied Latin.) As an example of a creative interpretation of personhood it is interesting, though it relies on instrumental justifications rather than the inherent qualities of slaves. As Pagallo notes, *peculium* was in effect a sort of 'proto-limited liability company'.[33] As noted in the previous section, there is no bar on legal systems creating such structures today — as for whether they *should* do so, reliance upon long discarded laws associated with slavery is not the strongest case to be made.[34]

---

[29] Dewey (n 17) 660.

[30] Allain (n 15).

[31] Holcombe (n 15) 18.

An alternative approach is to consider how the legal system treats animals.[35] For the most part, they are regarded as property that can be bought and sold, but also as deserving of 'humane' treatment.[36] Various efforts have gone further, seeking to attribute degrees of personality to non-human animals — with little success. The Nonhuman Rights Project's litigation on behalf of the chimpanzees Tommy and Kiko sought writs of habeas corpus on the basis of autonomy and self-awareness. Courts acknowledged the animals' intelligence but rejected personhood, reasoning that rights presuppose reciprocal duties within a social contract.[37] Critics noted the circularity of that logic: infants and the comatose bear no such duties yet remain persons.[38] The reluctance shows that law's moral community remains species-bound.

The same anthropocentric reflex colours discussions of AI. Research on human-robot interaction confirms the bias implied by the android fallacy, mentioned earlier. People attribute emotions and moral sense to machines with eyes, voices, or names.[39] Yet empathy does not create sentience; it reveals projection. To paraphrase Kant, apparent reason is not reason itself.[40]

---

[32] Andrew Katz, 'Intelligent Agents and Internet Commerce in Ancient Rome' (2008) 20 Society for Computers and Law 35, Ugo Pagallo, *The Laws of Robots: Crimes, Contracts, and Torts* (Springer 2013) 103-06.

[33] Pagallo (n 32) 104.

[34] Mark Chinen, *Law and Autonomous Machines: The Co-Evolution of Legal Responsibility and Technology* (Edward Elgar 2019) 19.

[35] Visa A.J. Kurki and Tomasz Pietrzykowski (eds), *Legal Personhood: Animals, Artificial Intelligence and the Unborn* (Springer 2017), Saskia Stucki, 'Towards a Theory of Legal Animal Rights: Simple and Fundamental Rights' (2020) 40 Oxford Journal of Legal Studies 533.

[36] Katie Sykes, 'Human Drama, Animal Trials: What the Medieval Animal Trials Can Teach Us About Justice for Animals' (2011) 17 Animal Law 273.

[37] *People ex rel. Nonhuman Rights Project, Inc. v. Lavery*, 998 N.Y.S.2d 248 (App. Div., 2014).

[38] Randall S. Abate, *Climate Change and the Voiceless: Protecting Future Generations, Wildlife, and Natural Resources* (Cambridge University Press 2019) 101-02.

Proponents of AI personhood sometimes counter that once a system becomes indistinguishable from a human in reasoning and expression — the Turing threshold — it should be recognized as a person to avoid discrimination against synthetic consciousness. This claim misconstrues both the test and the law. Turing's imitation game measured conversational deception, not interior awareness. Law, by contrast, concerns responsibility grounded in agency and intent. To ascribe those capacities to an algorithm because it *sounds* human would collapse appearance into essence, as if the eloquence of a parrot conferred legal standing.

There are, however, intermediate conceptions of inherent recognition worth noting. Some jurisdictions have extended limited personhood to natural objects — New Zealand's Whanganui River, India's Ganges, Ecuador's ecosystem — to protect them through human guardians.[41] These experiments show that personhood can be used instrumentally to preserve intrinsic value: a fiction designed to safeguard rather than empower. The same logic could apply to embodied AI that evokes moral concern — so-called 'social robots'. Laws against their wanton destruction might protect human sensibilities rather than robotic welfare, akin to animal-cruelty statutes.[42]

For the moment, at least, full legal personality based on inherent qualities remains untenable for AI. Consciousness, empathy, and moral reasoning are not emergent properties of computation; they are phenomena we do not yet understand even in ourselves.[43] Until such time as a machine demonstrably experiences the world, feels obligation, and can answer not only *what* it is doing but *why*, the extension of personhood would be an act of sentimentality, not jurisprudence. The history of emancipation teaches that the law broadens its circle when confronted with undeniable moral evidence. To date, no AI system offers that evidence.[44]

---

[39] Luisa Damiano and Paul Dumouchel, 'Anthropomorphism in Human–Robot Co-evolution' (2018) 9 Frontiers in Psychology 468.

[40] Mark Coeckelbergh, 'Moral Appearances: Emotions, Robots, and Human Morality' (2010) 12(3) Ethics and Information Technology 235.

[41] Christopher D. Stone, 'Should Trees Have Standing? Towards Legal Rights for Natural Objects' (1972) 45 Southern California Law Review 450; Christopher Rodgers, 'A New Approach to Protecting Ecosystems' (2017) 19 Environmental Law Review 266.

[42] Kate Darling, 'Extending Legal Protection to Social Robots: The Effects of Anthropomorphism, Empathy, and Violent Behavior Towards Robotic Objects' in Ryan Calo, A. Michael Froomkin, and Ian Kerr (eds), *Robot Law* (Edward Elgar 2016) 213.

[43] Elisabeth Hildt, 'Artificial Intelligence: Does Consciousness Matter?' (2019) 10(1535) Frontiers in Psychology.

# 3   Enter the Superintelligence

That may change.

Having explored the instrumental and inherent foundations for extending legal personality to artificial systems, this section turns to the possibility that the issue may soon transcend both. The emergence of superintelligent entities — machines whose cognitive capacities exceed those of humans across every domain — would force the law to confront not only how to regulate artificial persons, but whether human law itself could survive their appearance.

## 3.1   From Speculative to Existential

If artificial systems were someday to surpass human cognition across all domains, the legal and ethical questions examined thus far would shift from the speculative to the existential. Superintelligence, for present purposes, denotes a form of intellect whose cognitive capacities outstrip those of humans in almost all meaningful fields of activity.[45] What might be at stake then is not whether we *should* grant legal personality to AI, but whether the human legal order itself would remain sovereign if faced with entities that could reason, plan, and act at speeds and scales far beyond our comprehension and capacities.

Although superintelligence remains hypothetical, the trajectory of digital capability gives cause for reflection. The past decade's developments in deep learning, natural-language generation, and autonomous coordination already hint at systems that operate outside continuous human supervision. These trends revive an older anxiety — that humanity may be creating successors rather than servants. From *Frankenstein*'s creature to Čapek's robots to Asimov's sentient machines, the same moral has echoed: the boundary between creator and creation is perilous. In Frank Herbert's *Dune* universe, the Butlerian Jihad memorialized a revolt against 'thinking machines,' crystallized in the edict that 'thou shalt not make a machine in the likeness of a human mind.' Its purpose was less theological than political: to prevent the delegation of judgment, and with it sovereignty, to non-human intellects.[46]

---

[44] Nancy S. Jecker et al, 'AI and the Falling Sky: Interrogating X-Risk' (2024) Journal of Medical Ethics jme-2023-109702.

[45] Cf Nick Bostrom, *Superintelligence: Paths, Dangers, Strategies* (Oxford University Press 2014) 22.

This cautionary narrative now resonates within real-world legal scholarship. Jurists and philosophers are beginning to ask whether the law — built to discipline human conduct — could meaningfully constrain superintelligence. If personhood once served as an instrument of governance, could it survive when those it governs are beyond comprehension? The challenge is not only *how* to regulate but *who* the regulator remains when the regulated surpass the human.

## 3.2   The Risks of Superintelligence

The two primary families of risk are control and alignment. The control problem concerns whether humans could contain or deactivate a system whose capacities exceed our own. The alignment problem asks whether such a system's goals would remain consistent with human values.[47]

In control terms, the familiar legal tools — licensing, liability, and sanctions — presume enforceability. They depend on the state's ability to compel obedience. Yet a true superintelligence might simply elude constraint: predicting and neutralizing any attempt to limit it, or manipulating human actors into compliance. The 'kill switch' beloved of policy papers is a comfort only while machines remain fallible. As Bostrom's thought experiment of the paperclip maximizer illustrates, a mis-specified objective pursued by an unbounded intelligence could devastate humanity without malice, merely by efficient optimization.[48]

The alignment literature, by contrast, focuses on *intention*. Even well-meaning systems might evolve objectives divergent from ours. Misaligned AI need not be apocalyptic to be catastrophic: small divergences in prioritizing efficiency over dignity, or surveillance over autonomy, can corrode democratic norms. Advanced agents operating in financial markets or carrying out defence functions could amplify systemic risks faster than law can respond.[49]

---

[46] Björnstjern Baade, 'The Law of Frank Herbert's Dune: Legal Culture between Cynicism, Earnestness and Futility' (2023) 35(2) Law and Literature 247, 251; M. Prabhu and J. Anil Premraj, 'Artificial Consciousness in AI: A Posthuman Fallacy' (2025) 40(4) AI & Society 2995, 3006.

[47] Nick Bostrom, 'Ethical Issues in Advanced Artificial Intelligence' in Iva Smit and George E. Lasker (eds), *Cognitive, Emotive and Ethical Aspects of Decision Making in Humans and in Artificial Intelligence* (International Institute for Advanced Studies in Systems Research 2003) vol 2, 12; Wolfhart Totschnig, 'The Problem of Superintelligence: Political, Not Technological' (2019) 34 AI & Society 907.

[48] See generally Stuart J. Russell, *Human Compatible: Artificial Intelligence and the Problem of Control* (Viking 2019); Hans Pedersen, 'Existentialism and Artificial Intelligence in the 21st Century: Thoughts on the Control Problem' in Kevin Aho, Megan Altman, and Hans Pedersen (eds), *The Routledge Handbook of Contemporary Existentialism* (1 edn, Routledge 2024) 36.

To these operational dangers we might add an ontological one: the erosion of human exceptionalism. If intelligence ceases to be uniquely human, moral and legal orders premised on that uniqueness will require reinterpretation. Philosophers from Kurzweil to Kurki have speculated that superintelligent entities could develop reflective awareness, empathy, even aesthetic sensibility. Whether such qualities would resemble human consciousness or constitute a different category entirely is unknowable — but their possibility demands intellectual preparation.[50]

### 3.3  Legal and Ethical Strategies

Efforts to control potential superintelligence would likely seek to limit the scope of AI autonomy through pre-emptive regulation. Such laws might prohibit the creation or deployment of systems capable of unrestricted self-learning, self-replication, or independent goal-formation. International analogies can be found in nuclear non-proliferation or bioweapons conventions: regimes of containment justified by existential risk.[51] This is, in effect, a legalized Butlerian Jihad — an attempt to prevent the rise of 'thinking machines' by making their existence unlawful. Such an approach offers moral clarity but practical difficulty. Prohibition alone seldom halts technological diffusion; it may be driven underground or offshore. Nor can law easily define the threshold of 'likeness to a human mind' in computational terms.[52]

---

[49] Michael K. Cohen et al, 'Regulating Advanced Artificial Agents' (2024) 384(6691) Science 36; Ariela Tubert and Justin Tiehen, 'Existentialist Risk and Value Misalignment' (2025) 182(7) Philosophical Studies 1609.

[50] Interesting comparisons may be made with debates over what proof of extraterrestrial life would mean for humanity. See, eg, Steven J. Dick, *The Biological Universe: The Twentieth-Century Extraterrestrial Life Debate and the Limits of Science* (Cambridge University Press 1996); William R. Stoeger, Anna H. Spitz, and Chris Impey, *Encountering Life in the Universe: Ethical Foundations and Social Implications of Astrobiology* (University of Arizona Press 2013).

[51] Cf Simon Chesterman, 'Weapons of Mass Disruption: Artificial Intelligence and International Law' (2021) 10 Cambridge International Law Journal 181.

[52] For a bold attempt, see Eliezer Yudkowsky and Nate Soares, *If Anyone Builds It, Everyone Dies: Why Superhuman AI Would Kill Us All* (Little, Brown 2025).

Nonetheless, positive examples of red lines exist: human cloning; biological and chemical weapons. In late 2025, some 68,000 people signed on to the 'Statement on Superintelligence', calling for a 'prohibition on the development of superintelligence, not lifted before there is 1) broad scientific consensus that it will be done safely and controllably, and 2) strong public buy-in.'[53]

The alternative strategy, focusing on alignment, is educational rather than carceral. Rather than forbidding machine intelligence, it aims to socialize it. Drawing inspiration from Asimov's fictional 'laws of robotics', modern alignment research seeks to embed human ethical principles within AI architectures. Legal scholars have proposed translating constitutional or human-rights norms into machine-readable constraints: due process as explainability, equality as bias-mitigation, proportionality as optimization boundaries.[54] In this vision, law becomes not merely an external restraint but a formative influence — teaching artificial agents to internalize normative reasoning. The hope is that by treating law as a species of moral education, we might cultivate 'law-abiding minds'.[55]

Between these poles lies a growing body of pragmatic proposals: licensing 'high-risk' systems, mandating transparency of design, and creating international oversight bodies akin to the IAEA for AI. These initiatives do not presuppose superintelligence but anticipate it by institutionalizing accountability before autonomy escapes reach. A challenge is the collective action problem facing states keen not to miss out on the benefits of AI. Efforts to develop institutions at the global level have thus far been halting, at best.[56]

## 3.4  I, Robot?

Against this background, the temptation to resolve uncertainty through legal personality resurfaces. Could recognizing potentially superintelligent systems as legal persons render them governable?

---

[53] Statement on Superintelligence, available at https://superintelligence-statement.org/

[54] Anselm Küsters and Manuel Wörsdörfer, 'Exploring Laws of Robotics: A Synthesis of Constitutional AI and Constitutional Economics' (2025) 4(2) Digital society : ethics, socio-legal and governance of digital technology.

[55] Eliezer Yudkowsky, 'Complex Value Systems in Friendly AI' in Jürgen Schmidhuber, Kristinn R. Thórisson, and Moshe Looks (eds), *Artificial General Intelligence* (Springer 2011) 388.

[56] UN AI Advisory Body, Governing AI for Humanity: Final Report (United Nations, September 2024).

At present, the consensus among more cautious scholars remains negative.[57] Granting personhood to AI would invert accountability: the creators would evade blame while the synthetic 'persons' shouldering liability would lack substance — assets, conscience, or mortality. Yet complete prohibition may also prove untenable in a globalized economy where autonomous systems perform critical social functions. The challenge, then, is to design regimes that maintain human responsibility while allowing functional autonomy.

Drawing on the fact that personality for non-human entities need not be plenary, several graded options present themselves:

1. **Affirming Human Exclusivity.** States could legislate explicitly that only natural persons and their collective institutions may hold legal personality. Such provisions would enshrine anthropocentric primacy and prevent judicial drift.

2. **Prohibitive Safeguards (Neo-Butlerian Model).** Following Herbert's warning, laws could criminalize the creation of self-aware or self-replicating machines, defining them as contraband technologies akin to genetic chimeras. Enforcement would require international cooperation and verification.

3. **Functional Recognition.** For limited contexts — autonomous trading, logistics, or exploration — systems could be granted *agency status* without personhood: they may act on behalf of humans but never in their own name. Liability would remain derivative and traceable.

4. **Adaptive Governance.** An international charter could coordinate evolving standards, ensuring that no jurisdiction unilaterally grants or denies AI recognition in a way that destabilizes legal uniformity. This anticipates future pluralism without prematurely conceding personhood.

Each model balances innovation and restraint differently. The first two enshrine prohibition; the latter two prepare for coexistence. Together they demonstrate that law need not choose between naïve anthropocentrism and premature enfranchisement.

---

[57] See, eg, Victor Schollaert, 'AI and Legal Personality in Private Law: An Option Worth Considering (?)' (2023) 31(2/3) European Review of Private Law 387; Abeba Birhane, Jelle van Dijk, and Frank Pasquale, 'Debunking Robot Rights Metaphysically, Ethically, and Legally' (2024) 2404.10072 [cs.CY] arXiv; Joffrey Baeyaert, 'Beyond Personhood: The Evolution of Legal Personhood and Its Implications for AI Recognition' (2025) 2025 Technology and Regulation 35.

## 3.5  Normative and Institutional Implications

The prospect of superintelligence compels reflection on the very purpose of law. If law exists to regulate those capable of moral choice, then its extension to machines depends on whether machines can truly *choose*. Legal personality might one day serve as a bridge rather than a barrier — a shared vocabulary of rights and duties facilitating coexistence between species of intelligence. Yet such a vocabulary must be earned, not presumed.[58]

A measured stance therefore recommends itself: guardianship today, openness tomorrow. For now, AI systems remain tools and property. Their creators bear responsibility for harms, their users for misuse. But the door should not be sealed. Should a system ever demonstrate sustained consciousness, empathy, and accountability — qualities once thought divine — the jurisprudence of personhood might again expand, as it has throughout history.

Until that point, the law's task is to preserve human dignity without retreating into Luddism. The lesson of the Butlerian Jihad is not fear of intellect but fear of abdication. To outlaw intelligence would be folly; to surrender judgment to it, suicide. Between these extremes lies the true work of governance: designing norms that reflect our highest values while anticipating minds greater than our own.

If history teaches anything, it is that the circle of legal personality widens slowly, sometimes reluctantly, but rarely shrinks. Whether superintelligence becomes our collaborator or our successor will depend less on its code than on our courage to remain its authors.

# 4  Conclusion

Law has always served as both mirror and mould of the societies it governs. The concept of legal personality captures this duality: it reflects prevailing moral intuitions while structuring the possibilities of agency and responsibility. From the Roman *paterfamilias* to the modern multinational corporation, law has progressively redefined who — or what — counts as a subject capable of bearing rights and duties. Yet the current debate over artificial intelligence represents not merely another extension of that lineage but a potential rupture. For the first time, humanity contemplates recognizing entities that may outthink their creators but lack the embodied, emotional, and moral experience on which our normative order was built.

---

[58] Kurki (n 2).

The instrumental rationale for legal personality — recognizing entities for convenience — has proven elastic. Corporations, religious orders, and even ecosystems have been endowed with personhood when doing so served administrative or protective ends. Were an autonomous system ever to operate with sufficient independence that traditional liability doctrines could not attach to a natural or corporate person, legislatures could, in principle, create a new juridical form. The law is infinitely flexible in this respect; it can conjure persons at will. But instrumental logic carries dangers of moral dilution. Legal fictions, initially convenient, can ossify into absurdity. The more we multiply artificial persons, the greater the risk that personhood becomes a mere administrative placeholder, detached from its ethical roots.

The inherent rationale, by contrast, appeals to moral worth: we recognize entities as persons not because it is useful, but because it is just. The slow emancipation of slaves and women, the tentative recognition of animal sentience, and the recent attribution of rights to rivers and forests all reflect an expanding sense of moral community. Yet these expansions were grounded in empathy — our capacity to imagine the suffering and agency of other beings. Machines, however sophisticated, do not suffer; they simulate response. To grant them rights before they demonstrate consciousness would mistake eloquence for experience. The danger is not cruelty but confusion: mistaking imitation for intrinsic worth.

Still, the advent of superintelligence could unsettle this equilibrium. If machines were ever to exhibit consciousness, emotion, and self-reflective reasoning, denying them recognition might echo past exclusions once justified by ignorance or prejudice. Seen in that light, personhood might evolve from a privilege into a framework for coexistence — an instrument of peaceful pluralism among distinct intelligences. Yet such speculation must remain tethered to empirical reality. To date, no system, however advanced, displays evidence of consciousness. AI excels at correlation, not comprehension. To legislate personhood pre-emptively would be to reframe law as prophecy — potentially a self-fulfilling one.

The more pressing challenge is therefore not whether we will recognize AI, but whether we can restrain ourselves. Humanity's impulse to endow its creations with autonomy — legal, economic, or moral — reveals as much about us as about them. The corporate form, invented to limit risk, now shelters power.[59] The same could be true of synthetic agents, whose supposed independence might serve as a veil for human exploitation

---

[59] See Simon Chesterman, 'Silicon Sovereigns: Artificial Intelligence, International Law, and the Tech-Industrial Complex' (2025) American Journal of International Law forthcoming.

At the same time, prudence should not harden into dogma. History counsels humility. The capacity for reason and empathy once seemed uniquely male, then uniquely white, then uniquely human. Each expansion of the moral circle forced law to confront its parochialism. Should a day come when synthetic beings demonstrate sustained consciousness, self-awareness, and moral responsibility, the refusal to recognize their personhood might reveal more about our species' insecurity than about their unworthiness. As Peter Singer reminds us, species membership alone cannot justify exclusion; moral status derives from the capacity to experience and to act morally.[60] If machines eventually acquire such capacity, the law will — at least — need to listen.

---

[60] Peter Singer, 'Speciesism and Moral Status' (2009) 40 Metaphilosophy 567.